%% file: main.tex
\documentclass[twocolumn, twocolappendix]{openjournal}
\usepackage{multirow}
\usepackage{xcolor}
\graphicspath{{./}{/}}
\usepackage{csquotes}
\usepackage{hyperref}
\hypersetup{colorlinks=true,linkcolor=blue,citecolor=blue,filecolor=blue,urlcolor=blue}
\usepackage{cleveref}
\usepackage{bm}		% Bold maths symbols, including upright Greek
\usepackage{orcidlink}
\input{macros.tex}
\shorttitle{The TEMPO Survey II}
\shortauthors{Soares-Furtado, et al.}
\begin{document}

\title{The TEMPO Survey II: Science Cases Leveraged from a Proposed 30-Day Time Domain Survey of the Orion Nebula with the Nancy Grace Roman Space Telescope}

\author{Melinda Soares-Furtado\,\orcidlink{0000-0001-7493-7419}$^{1,2}$}
\author{Mary Anne Limbach\,\orcidlink{0000-0002-9521-9798}$^{3}$}
\author{Andrew Vanderburg\,\orcidlink{0000-0001-7246-5438}$^{2}$}
\author{John Bally\,\orcidlink{0000-0001-8135-6612}$^{4}$}
\author{Juliette Becker\,\orcidlink{0000-0002-7733-4522}$^{1}$}
\author{Anna L.~Rosen\,\orcidlink{0000-0003-4423-0660}$^{5,6}$}
\author{Luke G.~Bouma\,\orcidlink{0000-0002-0514-5538}$^{7}$}
\author{Johanna M.~Vos\,\orcidlink{0000-0003-0489-1528}$^{8}$}
\author{Steve B.~Howell\,\orcidlink{0000-0002-2532-2853}$^{9}$}
\author{Thomas G.~Beatty\,\orcidlink{0000-0002-9539-4203}$^{1}$}
\author{William M.~J.~Best\,\orcidlink{0000-0003-0562-1511}$^{10}$}
\author{Anne Marie Cody\,\orcidlink{0000-0002-3656-6706}$^{11}$}
\author{Adam Distler\,\orcidlink{0009-0006-4294-6760}$^{1}$}
\author{Elena D'Onghia\,\orcidlink{0000-0003-2676-8344}$^{1}$}
\author{Ren\'{e} Heller\,\orcidlink{0000-0002-9831-0984}$^{12}$}
\author{Brandon S.~Hensley\,\orcidlink{0000-0001-7449-4638}$^{13}$}
\author{Natalie R.~Hinkel\,\orcidlink{0000-0003-0595-5132}$^{14}$}
\author{Brian Jackson\,\orcidlink{0000-0002-9495-9700}$^{15}$}
\author{Marina Kounkel\,\orcidlink{0000-0002-5365-1267}$^{16}$}
\author{Adam Kraus\,\orcidlink{0000-0001-9811-568X}$^{10}$}
\author{Andrew W.~Mann\,\orcidlink{0000-0003-3654-1602}$^{17}$}
\author{Nicholas T.~Marston\,\orcidlink{0009-0007-1123-0038}$^{1}$}
\author{Massimo Robberto\,\orcidlink{0000-0002-9573-3199}$^{18,19}$}
\author{Joseph E.~Rodriguez\,\orcidlink{0000-0001-8812-0565}$^{20}$}
\author{Jason H.~Steffen\,\orcidlink{0000-0003-2202-3847}$^{21}$}
\author{Johanna K.~Teske\,\orcidlink{0009-0008-2801-5040}$^{22}$}
\author{Richard Townsend\,\orcidlink{0000-0002-2522-8605}$^{1}$}
\author{Ricardo Yarza\,\orcidlink{0000-0003-0381-1039}$^{23,24}$}
\author{Allison Youngblood\,\orcidlink{0000-0002-1176-3391}$^{25}$}

\affiliation{$^1$Department of Astronomy,  University of Wisconsin-Madison, 475 N.~Charter St., Madison, WI 53706, USA}
\affiliation{$^2$Department of Physics \& Kavli Institute for Astrophysics and Space Research, \\ Massachusetts Institute of Technology, Cambridge, MA 02139, USA}
\affiliation{$^3$Department of Astronomy, University of Michigan, Ann Arbor, MI 48109, USA}
\affiliation{$^4$Center for Astrophysics and Space Astronomy, \\
Astrophysical and Planetary Sciences Department,  University of Colorado, Boulder, CO 80389, USA}
\affiliation{$^5$Department of Astronomy, San Diego State University, San Diego, CA 92182, USA}
\affiliation{$^6$Computational Science Research Center, San Diego State University, San Diego, CA 92182, USA}
\affiliation{$^{7}$Department of Astronomy, MC 249-17, California Institute of Technology, Pasadena, CA 91125, USA}
\affiliation{$^{8}$School of Physics, Trinity College Dublin, The University of Dublin, Dublin 2, Ireland}
\affiliation{$^{9}$NASA Ames Research Center, Moffett Field, CA 94035, USA}
\affiliation{$^{10}$Department of Astronomy, The University of Texas at Austin, Austin, TX 78712, USA}
\affiliation{$^{11}$SETI Institute, 189 N.~Bernardo Ave \#200, Mountain View, CA 94043, USA}
\affiliation{$^{12}$Max Planck Institute for Solar System Research, Justus-von-Liebig-Weg 3, 37077 G\"{o}ttingen, Germany}
\affiliation{$^{13}$Jet Propulsion Laboratory, California Institute of Technology, 4800 Oak Grove Drive, Pasadena, CA 91109, USA}
\affiliation{$^{14}$Physics \& Astronomy Department, Louisiana State University, Baton Rouge, LA 70803, USA}
\affiliation{$^{15}$Department of Physics, Boise State University, Boise ID 83725-1570, USA}
\affiliation{$^{16}$Department of Physics \& Astronomy, Vanderbilt University, Nashville, TN 37235, USA}
\affiliation{$^{17}$Department of Physics \& Astronomy, The University of North Carolina at Chapel Hill, Chapel Hill, NC 27599-3255, USA}
\affiliation{$^{18}$Space Telescope Science Institute, 3700 San Martin Drive, Baltimore, MD 21218, USA}
\affiliation{$^{19}$Johns Hopkins University, 3400 N.~Charles St., Baltimore, MD 21218, USA}
\affiliation{$^{20}$Center for Data Intensive and Time Domain Astronomy, Department of Physics and Astronomy, Michigan State University, East Lansing, MI 48824, USA}
\affiliation{$^{21}$University of Nevada, Las Vegas \& Nevada Center for Astrophysics, 4505 S Maryland Pkwy, Las Vegas, NV 89154, USA}
\affiliation{$^{22}$Earth and Planets Laboratory, Carnegie Institution for Science, 5241 Broad Branch Rd NW, Washington, DC 20015, USA}
\affiliation{$^{23}$Department of Astronomy and Astrophysics, University of California, Santa Cruz, CA 95064, USA}
\affiliation{$^{24}$Texas Advanced Computing Center, University of Texas, Austin, TX 78759, USA}
\affiliation{$^{25}$Exoplanets and Stellar Astrophysics Laboratory, NASA Goddard Space Flight Center, Greenbelt, MD 20771, USA}

\email{Corresponding author: mmsoares@wisc.edu}

\begin{abstract}
The TEMPO (Transiting Exosatellites, Moons, and Planets in Orion) Survey is a proposed 30-day observational campaign using the Nancy Grace Roman Space Telescope.
By providing deep, high-resolution, short-cadence infrared photometry of a dynamic star-forming region, TEMPO will investigate the demographics of exosatellites orbiting free-floating planets and brown dwarfs --- a largely unexplored discovery space. 
Here, we present the simulated detection yields of three populations: extrasolar moon analogs orbiting free-floating planets, exosatellites orbiting brown dwarfs, and exoplanets orbiting young stars. 
Additionally, we outline a comprehensive range of anticipated scientific outcomes accompanying such a survey.
These science drivers include: obtaining observational constraints to test prevailing theories of moon, planet, and star formation; directly detecting widely separated exoplanets orbiting young stars; investigating the variability of young stars and brown dwarfs; constraining the low-mass end of the stellar initial mass function; constructing the distribution of dust in the Orion Nebula and mapping evolution in the near-infrared extinction law;  mapping emission features that trace the shocked gas in the region; constructing a dynamical map of Orion members using proper motions; and searching for extragalactic sources and transients via deep extragalactic observations reaching a limiting magnitude of $m_{\mathrm{AB}}=29.7$\,mag (F146 filter).
\end{abstract}

\keywords{Brown Dwarfs (185) --- Natural satellites (Extrasolar) (483) --- Exoplanets (498) --- Free floating planets (549) --- Surveys (1671) --- Young star clusters (1833)}

\maketitle
\section{Introduction}
\label{sec:intro}
Time-domain surveys have provided valuable insights into the diverse and evolving phenomena that emerge at the early stages of star and planet formation. 
Variability analyses of these regions have informed our knowledge on topics such as (sub)stellar mass accretion  \citep{1994ApJ...427..987B,2010MNRAS.406.1208D,2017A&A...599A..23V,2018AJ....156...71C,2020ApJ...891..182S}, rotational velocities of young brown dwarfs and pre-main-sequence stars \citep{2002ApJ...566L..29H,2007A&A...463.1081M}, and circumstellar disk evolution \citep{2002astro.ph.10520H}.
While optical transit surveys have yielded a wealth of knowledge about exoplanets orbiting main sequence (MS) stars \citep[e.g.,][]{2007PASP..119..923P,Collier2007,2013ApJ...767...95D,Sanchis-Ojeda2014,Winn2015,Zhu2018,2018AJ....155...89P,2019AJ....158...75H,Vanderspek2019,Uzsoy2021}, our understanding of exosatellite populations remains comparatively sparse \citep[e.g.,][]{Hippke2015,2017MNRAS.464.2687H,Teachey_2018, Vanderburg2021,2022AJ....164..252T,Limbach2024}. 

The forthcoming Nancy Grace Roman Space Telescope (formerly known as the Wide Field Infrared Survey Telescope; hereafter Roman) presents the potential to drastically advance our understanding of this untapped discovery space. 
More specifically, high-precision near-infrared (NIR) photometric transit searches among free-floating planets (FFPs; also known as ``isolated planetary-mass objects" or ``rogue" planets) and brown dwarfs is an emerging method to detect exosatellites\footnote{Following a precedent established in the literature, we use the term ``exosatellite" to refer to a companion orbiting an FFP, brown dwarf, or late M dwarf \citep{Kenworthy2015,Muirhead2019}.}  \citep[e.g.,][]{2017MNRAS.464.2687H,Tamburo2019,Tamburo2022,2022AJ....164..252T, Limbach2021, Limbach2023,Limbach2024}.
This technique is particularly effective when applied to young FFPs and brown dwarfs that are still self-luminous at $1-2\,\mathrm{\mu m}$.

In this paper, we present the broad scientific outcomes that will accompany the proposed Transiting Exosatellites, Moons, and Planets in Orion (TEMPO) Survey --- a 30-day time domain investigation of the Orion Nebula Cluster (ONC) and surrounding regions using Roman. 
\cite{Limbach2023} found that the ONC --- a young star cluster located in the Orion Molecular Cloud Complex --- serves as the optimal location for the TEMPO survey, due to its young age (1--3\,Myr; \citealt{Jeffries2007}), proximity ($390\pm2$\,pc; \citealt{2022A&A...657A.131M}) and dense stellar population (central density ${\sim}10^{4.7}$~stars/pc$^3$; \citealt{McCaughrean1994,Hillenbrand1998}). 
Roman's instrument specifications, particularly its large field of view and IR bandpass, will facilitate the concurrent monitoring of thousands of young, luminous targets. We refer the reader to Table~1 in \cite{Limbach2023} for a breakdown of the estimated number of monitored sources grouped by host type.
Moreover, this survey offers the sensitivity required to detect Titan-sized worlds transiting brown dwarfs and FFPs \citep{Limbach2023}.

In addition to probing exosatellite demographics, TEMPO presents an opportunity to investigate a wide range of astrophysical phenomena.
In this work, we outline the anticipated scientific outcomes enabled by such an investigation.
In Section~\ref{sec:design}, we present a summary of the TEMPO survey design specifications.
Section~\ref{sec:core} details the primary scientific objectives shaping the design of the proposed survey, as well as the simulated survey exosatellite yield.
In Section~\ref{sec:additional}, we present the broad scientific outcomes enabled by the TEMPO survey.
Subsequent subsections delve into specialized topics, including: probing the planetary-mass regime of the Initial Mass Function (Section~\ref{subsec:imf}); ONC membership analysis of substellar targets (Section~\ref{subsec:membership}); direct imaging detections of substellar targets (Section~\ref{subsec:imaging});  gyrochronological investigations 
(Section~\ref{subsec:rotation}); investigations of stellar formation, evolution, variability, and multiplicity (Section~\ref{subsec:stars}); the generation of regional dust maps (Section~\ref{subsec:dust}); investigations of the nature and origin of the Radcliffe Wave (Section~\ref{subsec:radcliffe}); and extragalactic source detection (Section~\ref{subsec:detectextraglactic}).
We conclude in Section~\ref{sec:conclusion} with final remarks.

\begin{table*}[thb]
    \centering
    \tabletypesize{\small} % Adjust as needed
    \begin{tabular}{lc}
    \hline
    Parameter & Value \\
    \hline
    Field of View & 0.28\,deg$^2$  \\
    Duration & Two observation windows of 15\,days each, separated by one year \\
    Spectral Band & F146, 0.93--2.00\,$\mathrm{\mu m}$ \\ 
    Exposure Time & 18\,s (6 reads at 3\,s each)\\
    Limiting Magnitude (1 month) & 29.7\,mag$_{\rm AB}$\\
    Photometric Precision (1 hr, F146 = 17\,mag$_{\rm AB}$) & 125\,ppm \\
    Photometric Precision (1 hr, F146 = 21\,mag$_{\rm AB}$) & 850\,ppm \\
    \hline
    \end{tabular}
    \label{tab:para}
    \caption{Observational parameters for the proposed TEMPO survey, utilizing the Roman WFI/F146 filter.}
\end{table*}

\section{TEMPO Survey Design Summary}
\label{sec:design}
Currently set to launch in 2027, Roman will be stationed at the second Lagrange point and will have a primary mission lifetime of five years \citep{Green2012,Spergel2013,Spergel2015}. 
Roman's main science goals encompass censuses of high redshift galaxies and a microlensing survey of the inner Milky Way.
Roman has two main scientific instruments: the Coronagraph Instrument and the Wide Field Instrument (WFI). 
The proposed TEMPO survey aims to harness the WFI---a high-spatial-resolution (0.11'' per pixel), wide-field ($0.28\,\mathrm{deg}^2$) camera.
%over $100\times$ that of the Hubble Space Telescope.
The instrument covers a wavelength range spanning $0.48$--$2.3$\,$\mathrm{\mu m}$, enabling NIR observations.
Drawing from the findings of \cite{Limbach2023} and \cite{Tamburo2022}, TEMPO will employ the WFI/F146 filter. 
Future work will explore the benefits of alternating between the WFI/F146 and WFI/F213 filters as a method of disentangling color-dependent host variability from achromatic transit signatures.

The observational parameters of the proposed TEMPO survey are provided in Table~\ref{tab:para}. 
This table includes the survey field of view, proposed duration, spectral band limits, exposure time, limiting magnitudes, and expected photometric precision. 
Comprehensive information on the survey's design specifications and potential noise sources can be found in \cite{Limbach2023}. 
Notably, TEMPO will be near-photon-noise-limited for ONC members of 0.2\,\Msun{} down to 1\,\MJ{}, corresponding magnitude range of $17$--$23$\,\ab{}.

\begin{figure*}[tbh!]
\centering
\includegraphics[width=0.8\textwidth]{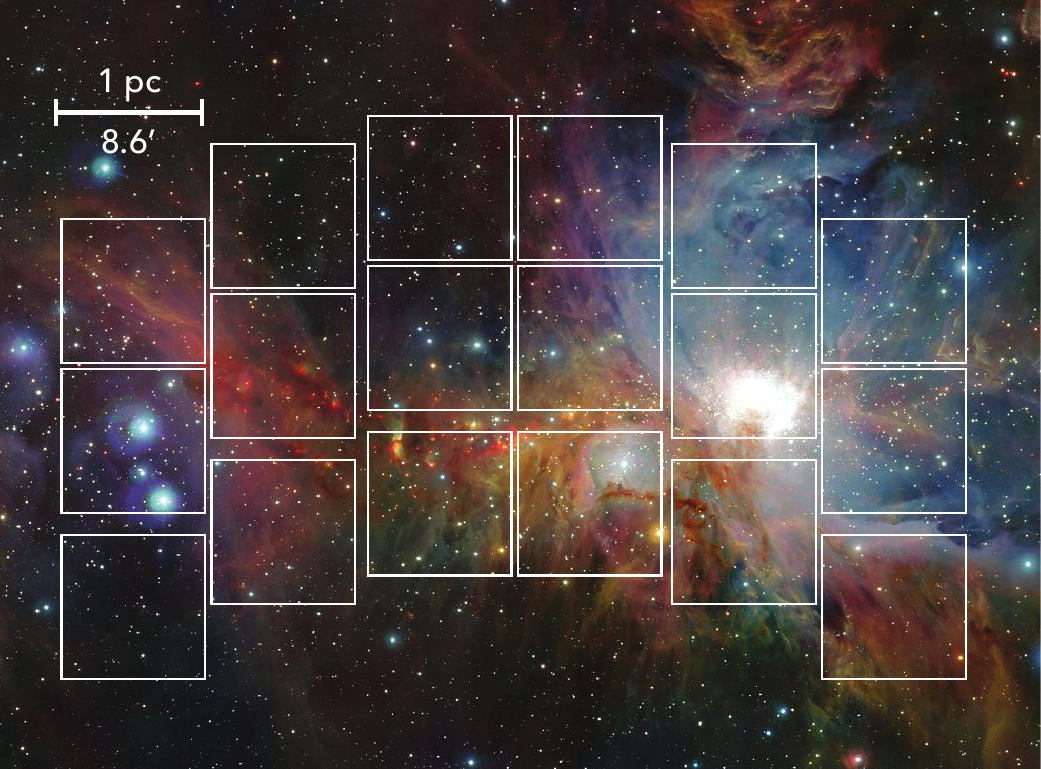}
%\caption{Field of view for the proposed TEMPO survey, which encompasses the Orion Nebula Cluster (1-3\,Myr). Credit: ESO/H.~Drass et al.}
\caption{Illustration of the proposed TEMPO survey's field of view, focusing on the Orion Nebula Cluster (ONC) which has an age range of 1--3\,Myr. The ONC is a region of intense star formation and offers a valuable window into the early stages of stellar and planetary evolution. Credit: ESO/H.~Drass et al.}
\label{Orion_FOV}
\end{figure*}

Prior successes, such as the Hubble Space Telescope (HST) F130N/F139M discovery of 320 brown dwarfs and 220 FFPs (with masses as low as 5\,\MJ{}) within a tiled FOV (13\,arcmin$^2$; \citealt{Gennaro2020}), suggest promising outcomes for TEMPO. Further, the TEMPO footprint offers more than twice the FOV obtained by prior tiled surveys of the ONC \citep{Drass2016,Robberto2020,Gennaro2020}. 
The TEMPO footprint is illustrated as the white outline in Figure~\ref{Orion_FOV}. The core of the ONC, known as the Trapezium cluster, is the brightest region in the figure. 
TEMPO couples Roman's larger FOV with an orientation optimized for maximum stellar density, allowing us to monitor $1.75\times$ the number of $17$--$21$\,\ab{} sources observed by \cite{Gennaro2020}.
Within the ONC alone, TEMPO will study 2,000 young stars, 560 brown dwarfs, and 400 FFPs \citep{Limbach2023}.
An additional 5,000 field stars of varying ages will also be monitored during the survey.

TEMPO can probe sources below the mass limit of the \cite{Gennaro2020} investigation, therefore, our estimate for the number of FFPs in our survey is likely to be on the conservative side. 
In addition, an investigation of the Trapezium Cluster using the James Webb Space Telescope (JWST) Near Infrared Camera (NIRCam) instrument reported the detection of 540 FFPs down to 0.6\,\MJ{} \citep{2023arXiv231001231P}. 
These sources were identified within an $11\arcmin{} \times 7.5$\arcmin{} ($1.2$\,pc $\times$ $0.8$\,pc) FOV, which is slightly larger than a single Roman tile. 
If we were to extrapolate from these findings, TEMPO will monitor as many as ${\sim}5,000$ FFPs, however, our reported yields remain conservative until there is further follow up of the JWST results.
%2.5*6*(1.75*220)

The TEMPO survey will observe the ONC and surrounding regions for 720 hours in two 15-day intervals separated by one year.
The expected photometric precisions range from 125\,ppm (for 17\,\ab{} stars) to 850\,ppm (for 21\,\ab{} stars). 
Continuous 18-s exposures will be employed to minimize detector saturation, providing unsaturated photometry for stars as bright as 17\,\ab{}.
Furthermore, we can obtain photometric measurements of the brighter, saturated stars by employing advanced analysis techniques, such as halo photometry \citep{White2017}. 
To assess cluster membership of dimmer sources, which fall below the faintness limit imposed by the \textit{Gaia} space observatory \citep{GAIADR3}, TEMPO proposes to measure kinematic (proper motion) by splitting the observation window into two 360-hour segments separated by one year. 
This science case is discussed in more detail in Section~\ref{subsubsec:kinematics}.
The one-year temporal gap in the observations also helps with vetting transit candidates, as transits are strictly periodic, while (sub)stellar variability is expected to evolve with time \citep{Limbach2023}.

\section{TEMPO Primary Science Outcomes}
\label{sec:core}

\begin{figure*}[tbh!]
\centering
\includegraphics[width=1\textwidth]{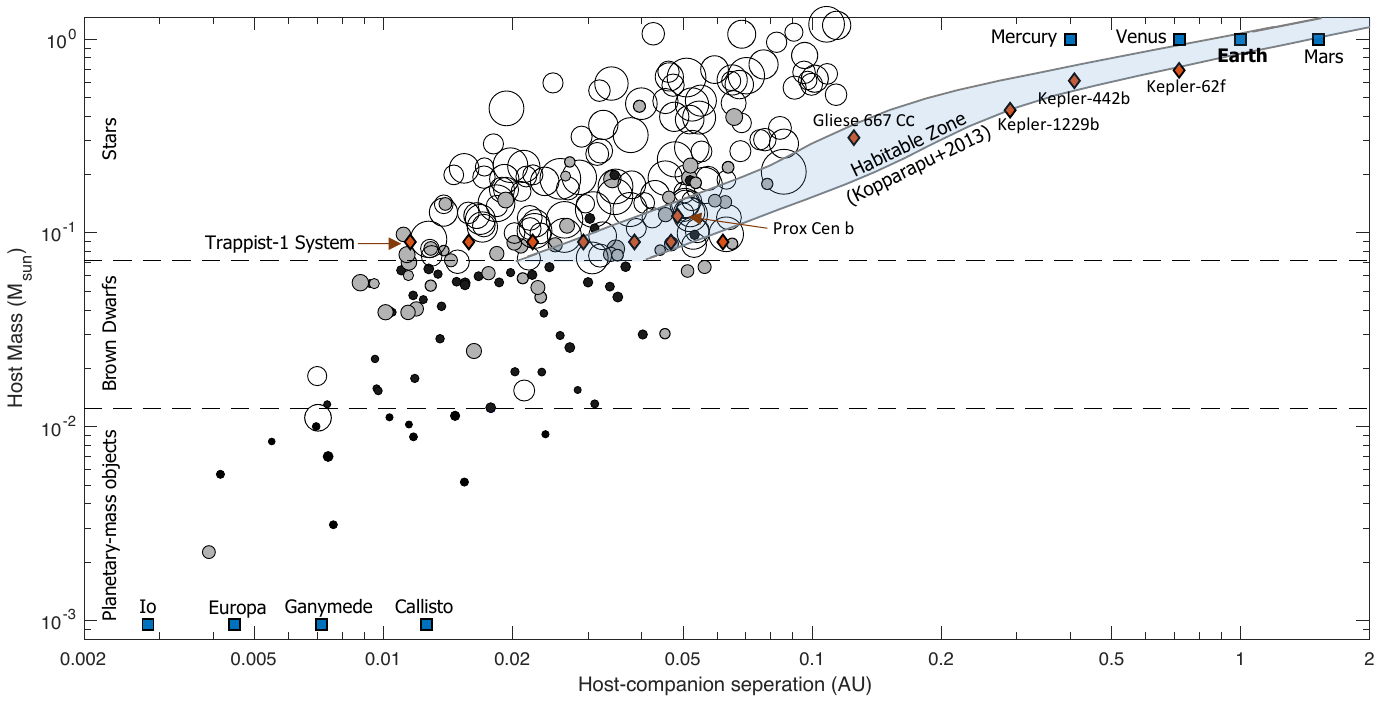}
\caption{Distribution of host-companion separation (AU) as a function of host mass (\Msun{}) for the simulated TEMPO survey detections (circles), Solar System moons and planets (blue squares) and select habitable exoplanets (red diamonds). Simulated TEMPO satellites and exoplanets: size of the circle corresponds to the satellite or planet mass. Open circles are gas giants ($>5$\,\ME{}), gray circles are Earths or super-Earths (1-5\,\ME{}) and black circles are super-Titans ($\mathrm{M_{Titan}}$-\ME{}). The light-blue region is the habitable zone for main sequence stars from \cite{Kopparapu2013}.}
\label{HZplot}
\end{figure*}

\subsection{Anticipated TEMPO Detection Yields}
\begin{figure*}[tbh!]
\centering
\includegraphics[width=0.59\textwidth]{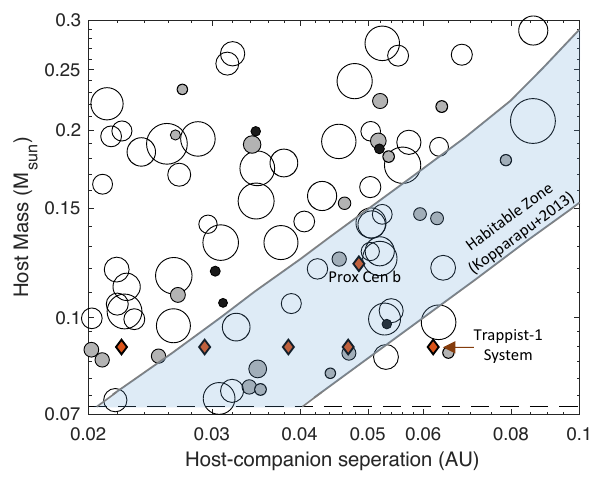}
\caption{Distribution of host-companion separation (AU) as a function of host mass (\Msun{}) for the simulated TEMPO survey detections (circles), Solar System moons and planets (blue squares), and select habitable exoplanets (red diamonds)---same as Figure~\ref{HZplot}, but zoomed in on the HZ. Simulated TEMPO exoplanets: circle size corresponds to exoplanet mass. Open circles are gas giants ($>5$\,\ME{}), gray circles are Earths or super-Earths (1-5\,\ME{}) and black circles are super-Titans ($\mathrm{M_{Titan}}$-\ME{}). 
The light-blue region is the HZ for main sequence stars from \cite{Kopparapu2013}. TEMPO is expected to detect approximately ten proto-habitable zone terrestrial ($<5$\,\ME{}) exoplanets around low-mass stars (mid to late M-dwarfs), similar to the Trappist-1 exoplanets and Proxima Centauri b.\\ }
\label{HZplotzoom}
\end{figure*}

In the first TEMPO paper, our team developed simulation code to calculate the expected yield from the TEMPO survey based on a wide variety of input parameters (satellite occurrence rates, radius-mass relations, ONC extinction, etc), and taking into account several possible instrumental configurations and measurement systematics (instrument spectral band, host variability, etc). In this section, we leverage that code to calculate the expected detections.
Figures~\ref{HZplot} and \ref{HZplotzoom} illustrate the estimated TEMPO survey exosatellite detection yields within the parameter space of orbital separation and host mass.
These figures incorporate two distinct sets of simulations: For the substellar hosts, the default parameters provided in Table~3 of \cite{Limbach2023} are used to calculate yield (settings used for this simulation are in bold font). For the stellar hosts, we leverage known occurrence rates around more massive stars (see distribution in Section~5.2.1 of \cite{Limbach2023}. 
The abrupt predicted increase in the detection of large ($>$sub-Neptune) exoplanets just above the stellar/sub-stellar boundary is an artifact that arises due to a change in the methods employed for yield calculations. This shift is due to the assumptions inherent in these calculations, which are subject to the various caveats and limitations detailed in \cite{Limbach2023}.

Also plotted are select Solar System moons and planets (blue squares), as well as several well-known habitable zone (HZ) exoplanets (red diamonds).
In both figures, we illustrate exosatellites in the following way: open circles indicate sub-Neptune to Jovian gas giant companions ($>5$\,\ME{}), gray circles indicate 1-5\,\ME{} companions, and black circles indicate super-Titan ($M_{\rm Titan}$-\ME{}) companions. 
The size of the circles scales with satellite mass.
The detection of companions of low-mass stars and high-mass brown dwarfs provide an opportunity to study extraordinary young counterparts to the currently known exoplanet population, while the detections of companions orbiting lower-mass brown dwarfs and planetary-mass hosts will constitute a critical discovery space, for which very little is currently known.  

The detection yields can be summarized as follows: 
\vspace*{-2.5mm}
\begin{itemize}
    \item[$\circ$] 12 Super-Titan ($\mathrm{M_{Titan}}$-\ME{}) exosatellites orbiting FFPs
    \vspace*{-2.5mm}
    \item[$\circ$] 2 $>1.0$\,\ME{} exosatellites orbiting FFPs
    \vspace*{-2.5mm}
    \item[$\circ$] 34 Super-Titan exosatellites orbiting brown dwarf hosts
    \vspace*{-2.5mm}
    \item[$\circ$] 20 $>1.0$\,\ME{} exosatellites orbiting brown dwarf hosts
    \vspace*{-2.5mm}
    \item[$\circ$] 21 Earth-sized ($<5$\,\ME{}) exoplanets orbiting young Orion stars
    \vspace*{-2.5mm}
    \item[$\circ$] 122 $>5.0$\,\ME{} exoplanets orbiting young Orion stars
    \vspace*{-2.5mm}
    \item[$\circ$] 20 exoplanets around older main sequence field stars 
\end{itemize}
 
\subsection{A Discussion of Critical Questions Addressed by TEMPO Transiting Companion Detections}
\label{subsec:exo}

While the initial TEMPO paper concentrated on creating the code that allowed us to estimate the expected yield of transiting companions (illustrated in the previous section), this manuscript will shift its focus to discussing critically important questions that can be addressed using the anticipated detected population.

\subsubsection{Detection of the First Population of Exosatellite Companions}

While the community has confirmed more than 5,500 exoplanets, we have yet to confirm the detection of a single exomoon/exosatellite. 
The literature reports approximately a dozen signals from candidate exomoons. These signals were identified through various methods, including gravitational microlensing \citep{Bennett_2014,Miyazaki_2018}, features observed in transit spectra \citep{Oza_2019,Gebek_2020}, gaps within circumplanetary rings \citep{Kenworthy_2015}, transit-timing variations (TTVs) accompanied by exomoon transits of the host star \citep{2018AJ....155...36T,Teachey_2018,2018A&A...617A..49R,Kreidberg_2019,2020AJ....159..142T}, TTVs alone \citep{Fox_2020,2020ApJ...900L..44K}, direct imaging \citep{Lazzoni_2020}, the detection of gas absorption possibly associated with an orbiting moon \citep{Ben_Jaffel_2014}, and exosatellites transiting FFPs \citep{Limbach2021} or more massive brown dwarfs \citep{2019sptz.prop14257M,2022AJ....164..252T}.

A critical next step for the astronomical community is the detection, confirmation, and characterization of exosatellite companions to substellar hosts. Of particular interest are the exosatellites of FFPs with masses similar to that of Jupiter. 
The TEMPO survey will enable the detection of exomoons/exosatellites in nearby, well-constrained environments, including the ONC and surrounding regions. This enhances our ability to interpret population demographics in these regions. The smallest exosatellites the TEMPO Survey is capable of detecting are $R \approx 0.35$\,\RE{} companions on short orbits (approximately 8\,hr) around 10-30\,\MJ{} hosts. 
This corresponds to predicted transit detections of exosatellites comparable in size to Titan, Ganymede, and Callisto, orbiting hosts with masses between 10-30\,\MJ{} \citep{Limbach2023}. Additionally, the survey will detect \enquote{Super-Titans} (exosatellites ranging from Titan to Earth-sized) transiting within all the Orion molecular clouds.
The proposed 720-hour observational window will allow for the detection of multiple companion transits at close orbital separations ($<0.1$\,AU). This outcome is crucial, as it provides an opportunity to distinguish between signatures of (sub)stellar variability and those of transiting companions \citep[e.g.,][]{Rizzuto2017,Mann2017,Mann2018,Rizzuto2020,Limbach2021,Ment2021}.

%The size of the smallest detectable transiting exosatellite is limited by the signal-to-noise of a given observation, which is dependent upon the transit duration, transit depth, photometric precision and the number of transits observed.
%\begin{figure}[tbh!]
%\centering
%\includegraphics[width=0.48\textwidth]{/Parameter_Space.pdf}
%\caption{The transit detection sensitivity space for a 3\,Myr $16$\,\MJ{} FFP host, using the modeling described in %\citet{Limbach2023} (pink-shaded region). We assume an SNR = 7 is required for exosatellite detection.
%White points denote major Solar System moons ($>0.03$\,\RE{}). 
%Red triangles denote exoplanets in Trappist-1.}
%\label{paramspace}
%\end{figure}

Identifying exosatellites bridges the knowledge gap between the population demographics of low-mass stellar systems, such as TRAPPIST-1, and the moon systems of the Solar System giant planets. 
Previous comparisons between compact \textit{Kepler} systems and the moons around giant planets in the Solar System by \citet{Kane13} have revealed similarities in system architectures, despite the host masses varying by several orders of magnitudes. 
TEMPO will enable comparative studies of companion architecture extending down to the planetary-mass host regime, several orders of magnitude lower than previously explored.

\subsubsection{Constraining Exosatellite Formation Theories}
The projected detection yields of the TEMPO survey vary significantly based on the selected assumptions in our simulations. By comparing the actual measured yield with the diverse outcomes predicted by our simulations, we can use the TEMPO results to determine which models align with observations. 
The TEMPO survey data will thus provide observational constraints to test prevailing theories of exomoon/exosatellite formation and evolution \citep[e.g.,][]{Canup2006,Heller2015,Barr2016,Miguel_2016,2018MNRAS.475.1347M,2018MNRAS.480.4355C,2020A&A...633A..93R,2020MNRAS.499.1023I,cilibrasi2020nbody,Nakajima2022}, which predict differing rates of companion occurrences and orbital periods. Additionally, it will provide constraints on other models related to young worlds, like the existence of H/He envelopes at an early stage in the formation of terrestrial worlds, as discussed in section \ref{subsubsec:envelopes}.

TEMPO observations may also enable constraints on the formation location of the substellar hosts. Numerical simulations indicate that it is possible for exomoons to remain gravitationally bound to their rogue hosts out to orbital separations of $0.1$\,AU (orbital period of $12$\,days) \citep{Rabago_2018,Hong_2018}. 
Finding more distantly separated satellite systems around free floating planets may offer clues as to whether these worlds formed in-situ or were ejected after forming in a circumstellar disk. 

In addition, it may be possible to understand how the composition of exomoons/exosatellites compares with the exoplanet, which could indicate formation scenarios for both. 
For example, there appears to be a correlation between elements within stars and their planets \citep[e.g.,][]{Bond10, Thiabaud15,Adibekyan2021}, a relationship that is particularly true for refractory elements such as Mg, Si, and Fe \citep{McDonough03, Unterborn23}. 
Given that stars, planets, and their satellites are all formed around the same time and from similar material, the composition of exomoons/exosatellites may mirror the host star. 
 
%, for which few constraints are available. 
%We refer the interested reader to \cite{Limbach2023} (particularly Figures 6 and 9) for a review of the assumption-dependent detection yields. 
%In short, the most optimistic yields are accompanied by the following assumptions: all exosatellites with masses greater than Mars possess H/He envelopes comparable to that of early-Earths \citep{Hayashi1979}, assuming an orbital period distrubtion similar to that of measured \textit{Kepler} mid-M-dwarf reported by \cite{2019AJ....158...75H}, and employing the exosatellite occurrence rates modeled by \citealt{Cilibrasi2021} where companion mass is scaled by the host mass, and incorporating a $0.01\%$ transit depth detection limit imposed by host variability.

%In this optimistic case, we expect TEMPO to detect approximately 22 FFP exomoons/exosatellites and 150 brown dwarf exoplanets/exosatellites. 
%In contrast, if $\sim$Earth-mass exosatellites do not still possess a H/He envelope, TEMPO is expected to detect only three FFP exomoons/satellites and eight brown dwarf exomoons/satellites. The extraordinary large range of survey yields illustrates how little is known about this population.
%A survey resulting in even the most limited yield estimates will provide an important opportunity to constrain this new discovery space. 

\subsubsection{Capture and Loss of H/He Envelopes}
\label{subsubsec:envelopes}

Theoretical models based on the current understanding of the known exoplanet population and the Solar System's terrestrial planets indicate that young rocky bodies with masses $>0.1$\,\ME{} (greater than the mass of Mars) are likely to possess gaseous envelopes composed of hydrogen and helium \citep{Hayashi1979}. 
The H/He envelope capture and dissipation models have far-reaching scientific implications, impacting the critical core mass for gas giant formation \citep{Hori2011}, the composition of the secondary atmosphere \citep{Misener2021}, and even the potential for the planet's future habitability \citep[e.g.,][]{Owen2016}.

H/He envelope capture and loss timescales are the largest driver in the expected yield of exosatellite and exoplanet yield for the TEMPO survey, as envelopes can significantly increase the sizes of these objects \citep[][e.g.,]{Lammer2020,Scherf2021,Wordsworth2021,Hayashi1979,Erkaev2014,Stokl2015,Stokl2016}.
\cite{Limbach2023} demonstrate that transit detection yields vary by nearly an order of magnitude depending on the implemented H/He envelope capture and loss model.

To date, there are no known systems that are young enough to constrain prevailing models of envelope capture and loss of terrestrial ($\sim$Earth-mass) worlds.
However, TEMPO will be capable of detecting a population of young, nearby exosatellites and exoplanets predicted to possess H/He envelopes \citep{Rogers2021}. Transit detection from the TEMPO survey alone will not be capable of distinguishing between low-mass proto-terrestrial worlds in possession of an envelope and more typical sub-Neptunes of similar radii (but much higher density).
In some cases, follow-up transmission spectroscopy via the HST and/or JWST is possible for targets brighter than F146~$\gtrsim 16$\,\ab{}.
%While it will be a challenge to disentangle the signatures of star spots or other inhomogeneities from the signatures of transiting exoplanets, it has been shown that stellar contamination in transmission spectra is most concerning in the ultraviolet
%and visual regime \citep{2018ApJ...853..122R,2019AJ....157...96R}.
%Additionally, it is possible to leverage the observed relation between a star's rotational variability amplitudes and the corresponding spot covering fraction \citep{2019AJ....157...96R}.

Such follow-up observations will offer the means to explore several significant questions that are presently challenging the field. 
This includes an opportunity to precisely constrain the exosatellite's mass, enabling differentiation between higher mass sub-Neptunes and proto-terrestrial worlds with H/He envelopes. 
Such observations will also constrain the atmospheric composition, including C/O ratios and overall metal abundances.
The presence of an extended envelope (i.e., large atmospheric scale height) enables such follow-up observations potentially down to Earth-mass companions. This will offer an unprecedented opportunity to examine the atmospheres of companions similar to early-Earth.

\begin{figure*}[tbh!]
\centering
\includegraphics[width=0.65\textwidth]{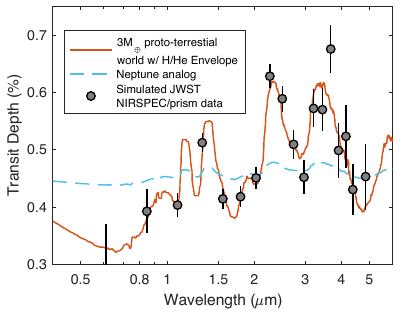}
\caption{The transmission spectra for a Neptune-analog (blue dashed line) and a 3\,M$_\Earth$ proto-terrestrial world with a H/He envelope (red line). Also shown is simulated JWST NIRSpec prism data of the proto-terrestrial world (error bars) assuming three hours of observations during transit(s) on a 16.1\,mag, 52\,M$_{\rm Jup}$ brown dwarf host. In this simulation, the extended envelope of the proto-terrestrial world is easily detected, however, cloud cover may impede detectability (see discussion in text).}
\label{JWSTSpecSim}
\end{figure*}

Figure~\ref{JWSTSpecSim} illustrates the simulated transmission spectra for a Neptune-analog with a clear atmosphere and a 3\,M$_\Earth$ proto-terrestrial world with a H/He envelope transiting a young 52\,M$_{\rm Jup}$ brown dwarf host. Both exosatellites have a radius of about 4\,$R_\Earth$ due to the presence of a very extended the H/He envelope on the proto-terrestrial world. Both exosatellites are modeled using clear atmospheres in \texttt{Exo-Transmit} \citep{2017PASP..129d4402K}. The simulated JWST data (black/gray error bars) of a 3\,M$_\Earth$ proto-terrestrial world illustrates its low-density atmosphere. The presence of thick clouds may make the atmosphere undetectable and/or degenerate with the Neptune-analog.

The simulation provided here is meant to illustrate the potential detectability of proto-terrestrial worlds discovered with TEMPO, however, we caution that further modeling beyond the scope of this paper will be required to fully and accurately model the atmosphere of a proto-terrestiral world. Here, we have made the simple assumption that \texttt{Exo-Transmit} can be used to model transmission spectra of any exoplanet with a H/He envelope. However, in low surface gravities, \texttt{Exo-Transmit} works under the assumption that the surface gravity is constant through the entire modeled portion of the atmosphere.  This is a fine assumption if the atmospheric thickness is much smaller than the planetary radius.
In the case of a low-$g$ extended envelope around a proto-terrestrial world, it will underestimate the transmission spectrum amplitude since, in this low surface gravity case, $g$ decreases significantly throughout the atmosphere. 
We recommend further modeling beyond the scope of this paper to explore this in more detail.

Probing the atmospheres of Earth-mass exoplanets will help to constrain the lower mass limit for the capture and loss of planetary envelopes.  
These data will provide an opportunity to test our understanding of terrestrial planet formation, as well as prevailing theories of Earth's formation \citep[e.g.,][]{Hayashi1979,Rogers_2011,Mordasini_2012,Stokl2016}.
TEMPO data may show that young exoplanets around low-mass stars have begun to lose their envelopes on timescales of a few Myr.
This will help differentiate between viable envelope dissipation mechanisms, such as photoevaporation via high-energy (XUV) radiation from the stellar host \citep[e.g.,][]{2020A&A...638A..52M}, core-powered mass-loss \citep[e.g.,][]{2018MNRAS.476..759G,2019MNRAS.487...24G, Rogers2021b}, and an initial boil-off stage \citep{Owen2016b,Owen2020}.

TEMPO data can be used to test theories of the underlying mechanism driving the observed ``radius gap" near 1.7\,\RE{} --- a theorized loss of H/He envelopes among less massive substellar bodies \citep[e.g.,][]{Lopez_2014,Fulton2017,Owen_2017,Owen2020,Misener2021}.
More specifically, it will be possible to probe 0.1-4\,\ME{} proto-terrestrial ONC exoplanets for the presence of H/He envelopes to determine if this radius gap is observed at these extremely early stages of evolution. 

\subsubsection{Habitable \& Pre-Habitable Worlds}
We refer to systems in the light-blue region in Figures~\ref{HZplot} and \ref{HZplotzoom} as ``proto-habitable zone" systems, as they indicate the location of the HZ once stellar hosts reach the main sequence \citep{Kopparapu2013}.
This is not to be confused with the location of the HZ at early timescales of 1-3\,Myr. 
In Figure~\ref{HZplotzoom}, we provide a zoom-in of ``proto-habitable zone" systems . 
TEMPO is expected to detect approximately ten proto-habitable zone terrestrial ($<5$\,\ME{}) exoplanets orbiting mid to late M-dwarfs. 
While a large fraction of the exosatellite detection yield extends to lower mass hosts, the majority of our expected TEMPO transit detections consist of systems similar to Proxima Cen b and Trappist-1 (but much younger).
Unlike stellar hosts, the brown dwarfs and FFPs will continuously cool, therefore never providing a long-lived stable HZ \citep{Barnes2013}. 
While we do not illustrate the HZ for FFP and brown dwarf hosts, it may be possible that detected exosatellites orbiting FFPs and brown dwarfs occupy a ``transient" HZ, which in some cases may persist over gigayear timescales \citep{Barnes2013}. 

HZ exoplanets orbiting main sequence stars have very low transit probabilities: 2\% for M-dwarf hosts and $<0.5$\% for sun-like hosts. 
While substellar worlds harbor a transient HZ, \cite{Limbach2021} showed that the transiting companions of young FFPs and brown dwarf hosts are far more likely to orbit within the HZ. 
This is the result of two factors: the HZ of young FFPs and brown dwarfs is very close to the host star, and because these hosts have large radii at young ages.
The typical transit probability of a HZ exosatellite in the TEMPO survey is approximately $10\%$.
TEMPO will offer an opportunity to help constrain the number of expected habitable exomoons and exoplanets, which in turn helps to quantify the total number of habitable worlds where we might expect to find life. 
%The HZ of brown dwarfs and FFPs is transient as the hosts cool \citep{Barnes2013}, but detected young HZ worlds can give us insight into ``Early-Earth".

The known population of detected exoplanets does not contain sources that are both  $<100$\,Myr \textit{and} $<1.8$\,\RE{}.
Therefore, there are no known proto-terrestrial worlds --- barring cases that may undergo substantial loss of extended atmospheres.
The dozens of young exoplanets expected to be detected by the proposed TEMPO survey will provide the first window into the properties of young terrestrial worlds. 
As discussed in Section~\ref{subsubsec:envelopes}, a young exoplanet's large scale height --- resulting from existing, but temporary, H/He envelopes --- will enable observations of follow-up transmission spectroscopy. 
If a proto-terrestrial world is detected around a sufficiently bright host star, it may be possible to characterize the proto-habitable-zone of ``primordial-Earth” exoplanets.
Such studies may offer a rare opportunity to place temporal constraints on the timescales for the formation of conditions favorable to the emergence of life. 
Under reasonable assumptions, we estimate the detection of about ten proto-habitable zone worlds transiting low-mass stars. 
Such systems will be young analogs to the 7\,Gyr Trappist-1 system, but at a much younger age ($\lesssim$3\,Myr).

Additionally, it has been proposed that some exomoons could harbor conditions that support liquid water --- and potentially life.  It is not a requirement for an exoplanet (or FFP) to be in the habitable zone in order for the exomoon/exosatellite to be conducive to life, since the moon could receive thermal radiation from the planet \citep[e.g.,][]{Scharf2006,Hinkel13, 2013AsBio..13...18H,Avila2021}.  
The tidal interactions among moons and planets can heat the moon interiors to produce liquid water \citep[e.g.,][]{Dobos2015,Roccetti2023} --- as is seen with Europa.  These necessary conditions can survive the planet's ejection \citep{Rabago_2018} and can persist for billions of years.  
\subsubsection{Constraining Exoplanet Formation Timescales}
%The detection and characterization of exoplanets orbiting 1-3\,Myr stars will help to constrain the timescales for exoplanet formation and disk dispersal. 
%Disks and exoplanet formation are intimately linked, as disks may play a critical role in the growth of exoplanets.

%ALMA observations provide evidence of planet formation while stellar hosts are still well within in the proto-stars stage of development \citep{Segura-Cox2020}.
%In some cases, observations indicate that planet formation occurs on timescales as short as $0.5$\,Myr \citep{Tychoniec2020}.
%In the case of young rocky planets, some theoretical studies support the formation over short timescales \citep[e.g.,][]{2023NatAs...7..330B}, while others claim that the process is likely to be significantly slower \citep{2016ApJ...817...90L}.

A long-standing conundrum is how short-period gaseous planets reach orbital periods of a few days or shorter. The discovery of the very first exoplanet around a Sun-like star, 51 Peg b \citep{Mayor1995}, inaugurated this open question, and it remains unresolved. 
Three main avenues for the origination of these puzzling planets have emerged \citep{2018ARA&A..56..175D}: in situ formation -- the gas giants formed in or near their current orbits; disk migration -- gravitational interactions between the nascent planets and their parental gas disks rapidly move the planets from where they form several AU from their host stars into their current orbits; and high-eccentricity tidal migration --- some flavor of gravitational perturbation excites their orbital eccentricities high enough that tidal interactions with their host stars circularize and shrink the orbits. 

No one explanation seems consistent with all available evidence, but a key diagnostic that may help distinguish between the different scenarios is the timescale required for migration. 
For in situ formation, short-period gaseous planets will be found in short-period orbits from the very earliest stages of planet formation, perhaps only $10^5-10^6\,{\rm yr}$ after the system originated \citep{2018ARA&A..56..175D}. 
It is likely that disk migration occurs early in the system's history, certainly before the protoplanetary disk is dissipated \citep{Kennedy2009}. The migration itself likely occurs faster than the formation of the plane --- perhaps several times $10^5\,{\rm yr}$, depending on the planet's mass and the disk properties \citep{2015A&A...574A..52D}. 
High-eccentricity tidal migration, by contrast, may act at any stellar age and should lead to a time-dependant hot Jupiter occurrence rate \citep{Miyazaki2023}. The ONC is sufficiently young (1-3\,Myr) that the detection of hot Jupiters with a similar occurrence rate to that which is seen around main sequence stars will provide evidence in favor of the first two theories, while a significantly lower occurrence rate in these young systems may place limits on the fraction of systems assembled by the last mechanism.

While current observational evidence (gleaned from measured orbital parameters such as orbital geometries and planetary eccentricities) suggests that tidal migration is the dominant pathway to form hot Jupiter systems \citep{Rodriguez2023, Zink2023}, it is clear that not all hot Jupiter systems form in this way \citep[e.g.,][]{Becker2015, Canas2019,Huang2020, Hord2022, Sha2023, Maciejewski2023}. The next step to understanding how hot Jupiters form is to refine the relative frequencies with which each of these three scenarios operate. TEMPO will provide a valuable constraint towards this goal because it will refine the hot Jupiter occurrence rate for the youngest stars, a diagnostic of exactly how important post-disk phase evolution is in shaping the hot Jupiter population we observe. 
TEMPO thus, through measurement of this occurrence rate, has the potential to provide a valuable additional constraint in the efforts to map out the relative importance of each hot Jupiter formation mechanism. The TEMPO constraint on occurrence rate in the young ONC, combined with constraints derived from surveys like TESS, will paint a more complete picture of hot Jupiter formation across all timescales. 

%help distinguish between these scenarios, complementing the ongoing efforts of surveys like TESS, in an effort to constrain the relative contributions of these formation pathways to the origins of short-period planets. 

\section{Additional Science Outcomes}
\label{sec:additional}
While the design of the survey is optimized for the detection of Orion exosatellites, TEMPO has the potential to make contributions to many other astrophysical contributions. 
We summarize those science cases in this section.

\subsection{Probing the Planetary-Mass Regime of the Initial Mass Function}
\label{subsec:imf}

Studying the planetary-mass regime of the Initial Mass Function (IMF) is crucial to refining theoretical models of star and planet formation.
Do planets frequently form like stars, via core collapse and turbulent fragmentation?
This formation channel will be an unusually low mass outcome of the normal star-forming process, as the mass of a planet is $100\times$ smaller than the average Jeans mass in star-forming clouds \citep{1997Sci...276.1836B}.
Yet, if planets are expected to form through the accretion of material in a circumstellar disk, this will indicate that the relatively large populations of FFPs observed in star-forming regions \citep[e.g.,][]{Gennaro2020, Robberto2020,2023arXiv231001231P} are likely ``rogue'' planets that have been ejected from their host stars. Is this supported by kinematic measurements? 

To probe this topic further, it is necessary to perform additional investigations of star-forming regions in the NIR.
Recent observations of the Orion Nebula and Trapezium Cluster using the Near Infrared Camera (NIRCam) on the NASA/ESA/CSA James Webb Space Telescope (JWST) revealed 540 planetary-mass objects with masses down to 0.6\,\MJ{} \citep{2023arXiv231001231P}.
Is it reasonable to expect such a large population of rogue planets in extremely young star-forming systems?
This discovery further challenges existing theories of planet formation, as 9\% of these detections consisted of planetary-mass objects (0.7-13\,\MJ{}) in wide, weakly-bound binaries. If these are rogue binary planets, how did such weakly bound system survive the ejection process? 

\begin{figure}[tbh!]
\centering
\includegraphics[width=0.48\textwidth]{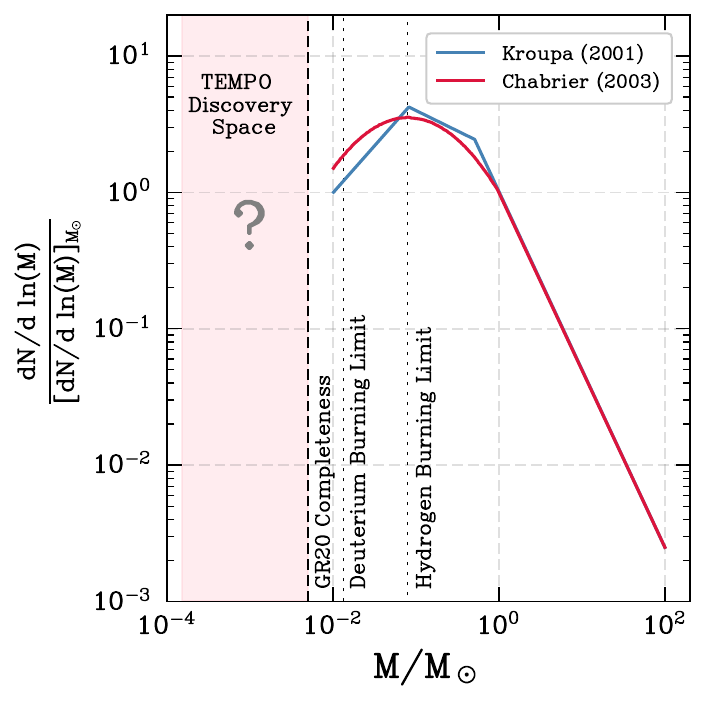}
\caption{Illustrated is the stellar initial mass function in terms of the number of stars formed per logarithmic interval in stellar mass (normalized to $M=1 M_{\odot}$). Overplotted is the IMF as described by \citealt{Kroupa2001} and \citealt{Chabrier2003}, as well as mass limits for the onset of hydrogen-burning and deuterium burning. 
The mass completeness limit of a low-mass IMF investigation of the ONC \cite{Gennaro2020} is also indicated. TEMPO will be sensitive to probing objects down to sub-Saturn masses (illustrated by the pink-hued panel).}
\label{fig:IMF}
\end{figure}

TEMPO has the ability to revisit this region, conducting a deeper, wide field, time series investigation. 
More specifically, with a limiting magnitude of 29.7\,\ab{} TEMPO can directly image FFPs with the following lower bound characteristics: $T_{\mathrm{eff}}\approx410$\,K and $R\approx1.4$\,\RJ{}, corresponding to a $50$\,\ME{} ($0.15$\,\MJ{}) exoplanet at an age of 1\,Myr \citep{Linder2019}.
The ability to directly image sub-Saturn mass FFPs will provide the lowest mass investigation of the IMF to date --- roughly one order of magnitude smaller than the FFPs that have been used as constraints in the past \citep{Gennaro2020}. 
Below several Jupiter masses, theory predicts that objects can no longer form via the normal star-formation process. 
Therefore, TEMPO will probe a mass region where the majority of the objects below ${\sim}3$\,\MJ{} are expected to be rogue exoplanets. While the number of objects found in this mass regime is poorly constrained, TEMPO's findings will have major implications for our understanding of the planet formation process.
Note that recent investigations have explored the low-mass end of the IMF of the ONC region down to 0.005\,$M_{\odot}$ \citep{Gennaro2020}.
The underexplored region of the substellar IMF is depicted as the pink-hued region in Figure~\ref{fig:IMF}.
The low-mass end of this pink-hued region is the lowest-mass FFP that TEMPO will be capable of observing ($50$\,\ME{}).

Building upon prior investigations, TEMPO will also have the opportunity to update the astrometric positions of detected FFPs, enabling kinematic investigations to compare the motion of FFPs with that of low-mass stellar members in the region. Kinematic measurements offer critical insights into the origins of these planets. 
While members of a stellar moving group generally follow a common spatial trajectory, the kinematics of an ejected planet should diverge from this shared trajectory.
Comparing the motion of FFPs to the motion of the cluster members will tell us the relative importance of the two formation pathways.

\subsubsection{Constraining FFP Formation Pathways With Exosatellite Demographics}
\label{subsubsec:kinematics}
The formation mechanism of FFPs also has important implications for the corresponding exosatellite demographics.
The dynamical encounters responsible for planetary ejection events have the potential to disrupt the orbital stability of the exosatellites, either causing the companion to collide with its host or supplying yet another rogue body to the surrounding environment as the exosatellite is stripped from the system \citep{Hong_2018}. 
Exosatellites that are weakly bound are particularly at risk of undergoing destabilization.

N-body investigations performed by \citealt{Hong_2018} and \citealt{Rabago_2018} indicate that FFPs with close-orbiting Io-like exosatellites are far more likely to survive dynamical scattering events.
These studies show that the FFP exosatellites that have survived dynamical scattering processes are more likely to be on short-period, eccentric orbits.
This offers predictions for the orbital configuration of exosatellites orbiting rogue planets, providing useful constraints on the viability of this potential FFP formation pathway.
This, in turn, offers further constraints on the planetary-mass regime of the IMF.

While it remains outside the scope of this work, \citealt{Rabago_2018} went on to show that it is possible for the Galilean-like exosatellites on eccentric orbits about FFPs to undergo significant tidal heating. 
This is an exciting result, which suggests that these exosatellites may be geologically active, with the potential to host subterranean oceans heated by a mechanism that could persist for billions of years. 

\subsection{Cluster Membership of Brown Dwarfs and Planetary-Mass Targets}
\label{subsec:membership}  

The current census of known members of the ONC is largely complete down to the spectral type of M5 \citep{mcbride2019}, thanks in large part to a number of spectroscopic surveys that are able to confirm the youth of stars, and the \textit{Gaia} mission which enabled identification of members via kinematic analysis. As these stars have only recently formed, not to mention their association with a dense cluster, their velocities tend to be similar to within a few km s$^{-1}$. Unfortunately, the census of cooler stars is limited \citep[e.g.,][]{defurio2021}: only a few of the brown dwarfs have been confirmed as members: since they are extremely faint, they are inaccessible to most surveys.
A deeper survey that can establish membership of brown dwarfs and free-floating planets is imperative in order to fully sample the IMF in such a notable young cluster. TEMPO will be a significant step towards providing this outcome.

Combining TEMPO astrometric measurements with existing HST and Keck II NIRC2 observations of the ONC \citep{kim2019} will significantly improve proper motion measurements of the red ONC stars that are too faint for astrometric analysis with \textit{Gaia}.
For reference, the faintness limit of the \textit{Gaia} survey is approximately 21\,\ab{}, while TEMPO will achieve a faintness limit of 29.7\,\ab{}. 
Such proper motions are required for a detailed analysis of stellar dynamics within the cluster \citep{getman2019}.

Typical proper motions of stars within Orion are approximately 2\,mas\,year$^{-1}$. Previous HST observations of young stars in the ONC achieved a positional accuracy of 2-3\,mas down to the limiting magnitude of 24.7\,mag in F775W band \citep{platais2020}. With a similar pixel scale, but observing at longer wavelengths, we expect to reach positional accuracy of approximately 5\,mas. This means that over a temporal baseline of one year, we expect to obtain precision in proper motions of approximately 5\,mas\,year$^{-1}$, or approximately 10\,km\,s$^{-1}$ at the distance of Orion. While such precision will not be sufficient for characterizing intracluster dynamics on its own (coupled with a large degree of extinction for stars behind the cloud), it will facilitate the identification of a number of low-mass members of the ONC from a significant number of field stars. It will also help to identify high-velocity runaways among already confirmed members \citep{mcbride2019}, such as, e.g, among dusty young stellar objects (YSOs) that have been primarily been detected in near-IR observations \citep{megeath2012}.

More importantly, TEMPO observations will act as a crucial anchor that will help to refine proper motions for these faint stars at subsequent observations of the cluster several years down the line. Indeed, combining TEMPO data with archival data, such as from HST observations that have been carried out from 2004--2015 \citep{kim2019,platais2020}, will double the temporal baseline, extending it to approximately 20 years, enabling a precision $<$0.2\,mas\,year$^{-1}$. 
While such a degree of precision is only possible for sources observed by both surveys (excluding sources fainter than 24.7\,mag and those outside the HST footprint), this underscores the importance of such legacy datasets, which TEMPO will be poised to become in the future. 

\subsection{Imaging Free Floating Planets and Widely-Separated Exoplanets}
\label{subsec:imaging}
Co-adding the full set of observations will permit the direct-image detection of FFPs with sub-Saturn masses (50\,\ME{} at 1\,Myr; \citealt{Linder2019}).
In addition, TEMPO will enable the investigation of a population of faint companions at wide separations using the advanced point spread function subtraction techniques based on Karhunen-Loeve Image Processing (KLIP) \citep{Pueyo2016}.
These techniques were recently applied to HST imaging data of the Orion Nebula \citep{Strampelli2021}.
Sub-Saturn mass planets at separations beyond $10$\,AU are largely an unexplored parameter space, yet characterizing these worlds remains critical to building a comprehensive framework of the overall architectures of planetary systems.

The detection of sub-Saturn planets requires the infrared sensitivity of an instrument like JWST or Roman --- albeit Roman enables the monitoring of a much larger field of view.
As described in Section~\ref{subsec:imf}, these observations also enable TEMPO to probe the low-mass-end of the IMF down to a largely untapped mass range. 

\subsection{Investigating Substellar Variability}
\label{subsec:rotation}
TEMPO will enable a large survey for substellar variability that will shed crucial light on the dependence of that variability on mass and spectral type in a young, coeval sample. Rotation periods for substellar objects at the age of the ONC are typically less than $\sim100$ hr \citep[e.g., see compilations by][]{Moore2019, Vos2022}, so multiple rotation periods will be covered in each TEMPO window, placing robust constraints on the rotational periods of the sample. This provides a unique opportunity to extend current gyrochronology studies into the substellar regime and to probe the angular momentum evolution of substellar objects as a function of mass at young ages. Planets within our Solar System show a clear trend between their rotational velocity and mass, and initial measurements of a small number of directly imaged exoplanets, FFPs, and brown dwarfs appear to follow this trend \citep{Snellen2014, Allers2016, Scholz2018}. Rotation period measurements by TEMPO survey will provide the first statistical sample to test this relation beyond our Solar System. 

TEMPO also enables a unique investigation of how light curve properties vary across spectral type and mass. For older FFPs and brown dwarfs, near-IR amplitudes depend on parameters including spectral type \citep{Radigan2014,Metchev2015}, surface gravity \citep{Metchev2015,Vos2022, Liu2023} and viewing angle \citep{Vos2017}. The small number of long-term monitoring studies of brown dwarfs and FFPs have shown that the light curves of variable objects evolve dramatically both on rotational and yearly timescales \citep[e.g.,][]{Apai2017, Zhou2022}. 
However, with relatively small sample sizes and samples that differ across ages, it is necessary that these trends are studied with a larger sample at constant age. TEMPO will enable investigations of the variability properties of substellar atmospheres in a uniform, coeval sample on a variety of timescales.

\subsection{Star Formation, Evolution, Variability, and Multiplicity}
\label{subsec:stars}  

\subsubsection{Investigating Star Formation}

The ONC and Orion Molecular clouds are among the most well-studied star formation regions in the Milky Way, exhibiting both low- and high-mass star formation. 
TEMPO Survey observations will permit a high spatial resolution investigation to map the dust distribution and examine stellar clustering.
The data can be leveraged to investigate the presence of disk dispersal via stellar feedback --- the injection of energy and momentum by young stars in the interstellar medium (ISM) --- and the presence of protoplanetary disks that may be actively undergoing planet formation.
TEMPO's spatial resolution (30\,AU) permits the mapping of the dense star-forming gas---including the presence of pillars, proplyds, and globules that are shaped by stellar feedback by young stars and are actively undergoing star and planet formation \citep{Hopkins2021a}. 
The presence of protoplanetary disks is supported by prior investigations, including a 3-year submillimeter investigation of the ONC, which found 42 individual protoplanetary disks (0.003-0.07\,\Msun{}) \citep{Mann2010}. 
Protoplanetary disks can be perturbed and/or truncated via stellar feedback from nearby young stars or dynamical effects in clustered environments like the ONC. For example, \cite{Otter2021} used ALMA data to investigate 127 protoplanetary discs in the ONC and surrounding regions. 
Of this sample, 72 were spatially resolved (3\,mm) and the disk sizes were smaller than expected --- likely a result of photoevaporation from nearby massive Trapezium stars.

Additionally, by mapping the density distribution of this region, we will determine the density PDF to compare to star formation models and determine how feedback shapes the ISM, drives turbulence, and may trigger star formation in star-forming environments \citep{Burkhart2018a, Rosen2020a, Menon2021a}.

\subsubsection{Rotation Rate Dispersion \& the Role of Disk Locking} 
Disk locking is an important mechanism driving the angular momentum evolution of stars during the first few million years ($\lesssim3$\,Myr) \citep[e.g.,][]{Edwards1993,Barnes2003}. 
Driven by magnetic interactions between young stars and their surrounding circumstellar disks, it is theorized that disk locking is responsible for the observed rotation rate dispersion among members within a young stellar cluster. 
At later stages, when circumstellar disks have been sufficiently dissipated, stars contracting onto the main sequence begin to spin more rapidly.
At this stage, the rotation rates of cluster sources are seen to converge, forming a tight period-mass-age relation \citep[e.g.,][]{Hartman2009,Meibom2015,Curtis2020}.

While precise circumstellar disk dispersal timescales are dependent on stellar mass, the surrounding radiation field, disk viscosity, and dust evolution, observations indicate that disk dispersal generally takes place within a few million years \citep[e.g.,][]{1993ARA&A..31..129L,Alexander2006}.  
The 3\,Myr ONC members targeted by TEMPO will offer an important opportunity to investigate the effects of disk locking on the rotational dispersion of an extremely young stellar population, while also probing the corresponding circumstellar disk dispersal timescales. 

\subsubsection{Investigating Pulsational Variability}
\label{subsubsec:var}
TEMPO will monitor approximately 10,000 targets for variability. 
This includes young stellar members in the ONC and surrounding Orion molecular clouds and approximately 5,000 field stars that fall within the TEMPO FOV, as shown in Figure~\ref{Orion_FOV}.  
Taking into account the photometric precision limits of the TEMPO survey (provided in Table~\ref{tab:para}), as well as the minimum integration time (18-s) and full 30-day observation window, the survey will be sensitive to the detection of a wide range of main sequence, subgiant, giant, and compact pulsational variables. 
This includes both gravity and pressure mode pulsational variables of the following classes: Solar-like pulsators, $\alpha$~Cyg, $\beta$~Cephei, Cepheids, $\delta$~Scuti, p-mode sdBVs (EC 14026 and V361~Hya), g-mode SdBVs (PG1716+426, lpsdBV), pulsating Be (SPBe), pre-MS pulsators (pulsating T Tauri stars, pulsating Herbig Ae/Be stars), $\gamma$~Doradus, GW~Vir, PG1716, roAp, RR Lyrae, SPB (53~Per), V777~Her, W~Vir, ZZ~Ceti, PNNV, and GW Vir.  

\subsubsection{Stellar Flares}
TEMPO Survey stars (both young and field stars) are expected to flare. Flares are sudden releases of energy that occurs when a star's magnetic field rearranges itself or reconnects \citep{Benz2010}, causing brightness variations on minutes-to-hours timescales characterized by a sharp rise and gradual decay phase (e.g., \citealt{TovarMendoza2022}). Flares are commonly observed in time-series data at all wavelengths \citep{Caramazza2007,Loyd2018, Macgregor2018,Howard2019}. Our knowledge of the energies and frequencies of stellar flares at infrared wavelengths is scarce \citep{Davenport2017, Howard2023}. 

More specifically, the study of how flaring impacts the infrared spectra for M-dwarfs is particularly necessary to measure accurate stellar abundances and exoplanet transmission spectra. In order to measure the composition of M-dwarf stars, which is difficult to accomplish on the ground due to the telluric interference of Earth's atmosphere, M-dwarfs must be observed in the infrared. One key issue affecting the determination of stellar abundances is the signature of flares filling elemental or molecular absorption lines. While this might be less important for early type (0-4) M-dwarfs which have a low flare occurrence rate \citep{Hilton10}, it is an important factor to understand for the more active late-type (5-9) M-dwarfs \citep{Kowalski09, Davenport12, Davenport16, Davenport19}. In addition, measuring the influence of flares on stellar spectra further into the infrared -- i.e. beyond the 2.5um threshold achieved by some of the most recent infrared surveys \citep{Schmidt12} -- will allow the community to determine signatures of flares beyond the lower order Paschen and Brackett lines. 
%Namely, it is not advisable to rely only on shorter wavelength flare indicators (such as the H$\alpha$ line at 0.656\,$\mu$m), which are not observable to many infrared instruments, or even to depend on previous or concurrent stellar observations. 

Flares are also a nuisance in the interpretation of exoplanet transmission spectra with JWST \citep{Lim2023}. Transit experiments generally rely on the assumption that stars are homogeneous and stable over time in order to isolate planetary spectral features from stellar features. However, stellar surface heterogeneity and time variability can imprint stellar spectral features in the resulting planetary spectra that are challenging to detect or remove \citep{Rackham2018}. Dedicated investigations into stellar flare properties at infrared wavelengths are required to mitigate their contamination of transit spectra \citep{Howard2023}.  

Understanding M-dwarf flares is critical not just for planetary detection, but for characterizing the planetary system and potential habitability, which is of importance to upcoming missions that are focusing on M-dwarf planets. Fortunately, Roman will make huge strides in this area \citep{TovarMendoza2023}, as Kepler/K2 did for stellar flares at visible wavelengths \citep{Shibayama2013, Hawley2014}. Probing the flaring statistics of M-dwarfs and coeval populations of stars with Roman will contribute to the growing body of research on the flaring evolution of cool stars \citep{Ilin2021}.

\subsubsection{Investigating Stellar Outflows}
The TEMPO survey will make significant contributions to studies of stellar jets and outflows. 
The depth of the TEMPO images will be unprecedented.  In the F146 filter, shocks emitting in the 1.27 and 1.46\,$\mathrm{\mu}$m [FeII] emission lines, and in the Paschen series of H-recombination lines will be clearly seen.  We will be able to use the F146 image, along with longer-wavelength images from JWST and the ground to remove stars and extended nebulosity.
The F146 filter traces ionized plasma from the Nebula and ionized by strong shocks and rendered visible in the Paschen series of H-recombination lines.   
Comparing TEMPO data to that of prior surveys --- including ground-based, HST, Gaia measurements, and JWST images --- will allow the measurement of proper motions of both stars and shocks with unprecedented precision.
Further, the inclusion of the F213 filter images will detect protostellar outflow shocks with unprecedented sensitivity.

Research has shown that the study of jets and outflows from young stars that the 2.12\,$\mathrm{\mu}$m line of H2 is the \textit{optimal} tracer of shock-waves propagating into molecular media \citep[e.g,][]{ReipurthBally2021,Bally2016}. 
JWST has shown that both the 2.12\,$\mathrm{\mu}$m v=1-0 S(1) line in the spectrum of molecular hydrogen is easily detected in both JWST's broad- and narrow-band filters in addition to the F212N narrow band filters. The Integral Shaped Filament  (ISF; the northern, densest portion of the Orion A cloud) located immediately behind the Orion Nebula extends approximately 30' South to 30' North of the Orion Nebula. 
The ISF is one of the richest regions of ongoing star formation, filled with jets and outflows best traced by the 2.12\,$\mathrm{\mu}$m line. TEMPO will enable investigations of the shock-waves propagating in this region, offering a new epoch to prior studies. 
There are hundreds of H2 shocks from dozens of jets and outflows already known in the Orion A ISF.  Summing the 30 days of data in the F213 filter will provide the deepest image of these flows.  Comparison with the exiting data will yield proper motions of the outflow shocks radiating in H2.   
Even if only 10\% of the time was devoted to the F213 filter, the resulting mosaic will be the deepest image of shock-excited (and fluorescent) molecular hydrogen emission in the ISF ever obtained.

\begin{figure}[tbh!]
\centering
\includegraphics[width=0.48\textwidth]{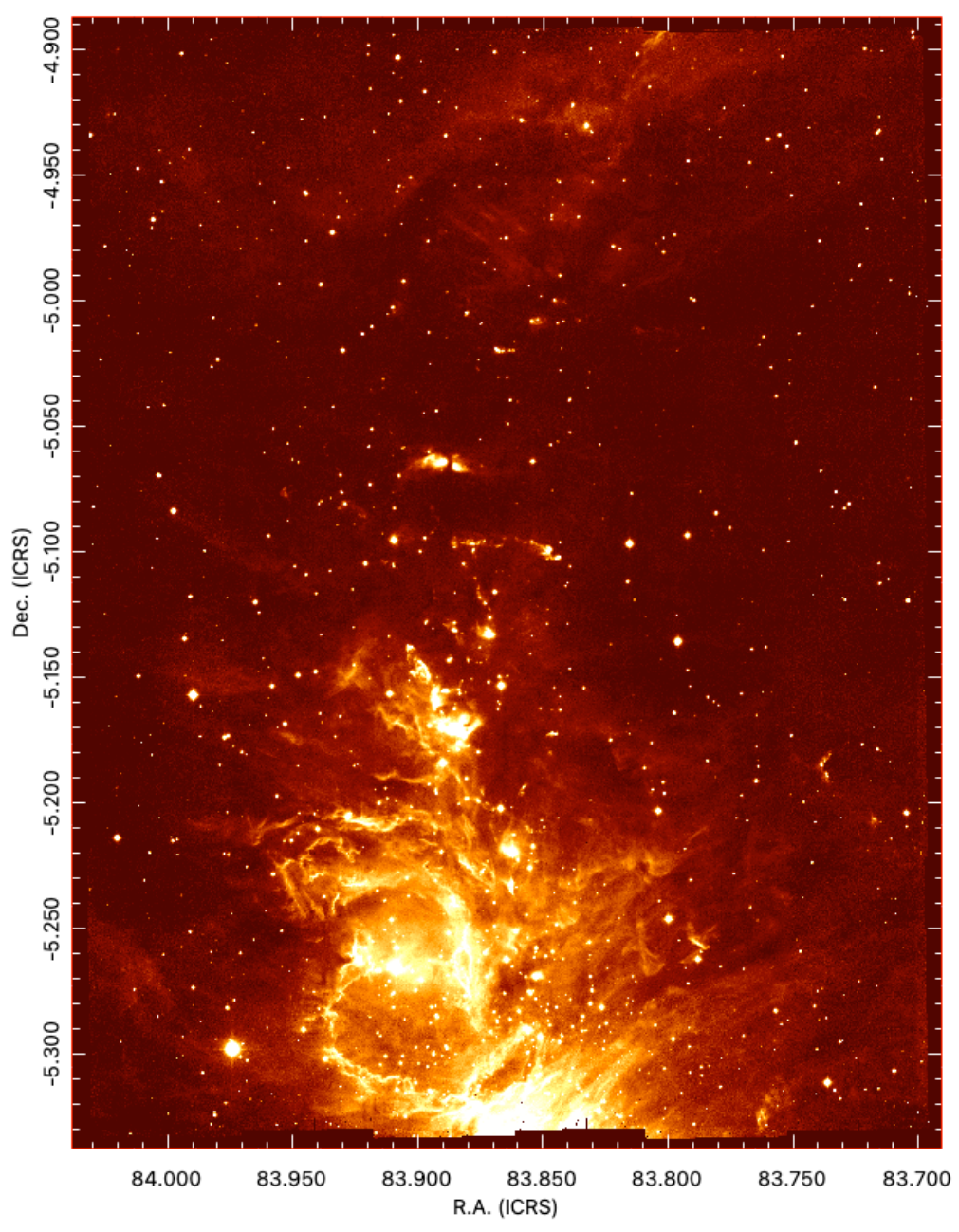}
\caption{Illustration of 2.12\,$\mathrm{\mu}$m H2 emission, tracing jets and outflows in the Northern field of the Integral Shaped Filament, immediately behind the Orion Nebula. These data were obtained by the Calar Astro 3.5-meter telescope in Spain \citep{Stanke2002}.}
\label{fig:outflowsn}
\end{figure}

\begin{figure}[tbh!]
\centering
\includegraphics[width=0.48\textwidth]{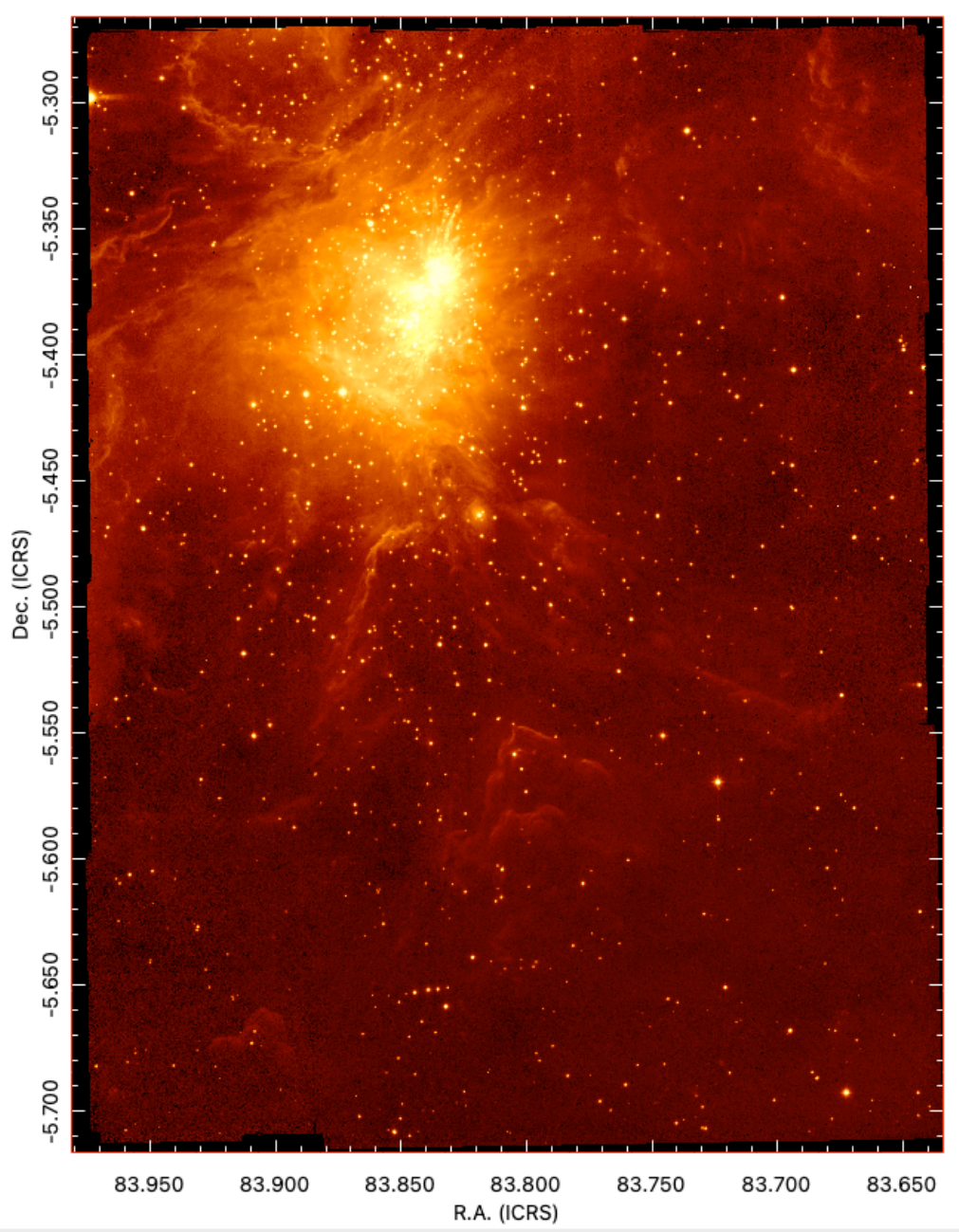}
\caption{Illustration of 2.12\,$\mathrm{\mu}$m H2 emission, tracing jets and outflows in the Northern field of the Integral Shaped Filament, immediately behind the Orion Nebula. These data were obtained by the Calar Astro 3.5-meter telescope in Spain \citep{Stanke2002}.}
\label{fig:outflowss}
\end{figure}

The jets and outflows traced out by 2.12\,$\mathrm{\mu}$m H2 emission can be seen in Figures~\ref{fig:outflowsn} and \ref{fig:outflowss}, which consists of a 1.2 degree field of the ISF, as observed by the Calar Astro 3.5-meter telescope in Spain \citep{Stanke2002}.
Both images contain significant overlap with the TEMPO field.
Figure~\ref{fig:outflowsn} illustrates the Northern field, while Figure~\ref{fig:outflowss} illustrates the Southern field. 
The Northern field shows the M43 cavity, and a host of molecular outflows and jets emerging from OMC2 and OMC3.  
The very bright emission in the Southern field is the Orion Nebula.   
This field contains the OMC1 and OMC4 regions.   

Additionally, the nearest group of currently forming massive stars (embedded in Orion OMC1, 0.1\,pc behind the Nebula) power the "Orion Fingers" of shock-excited H2 material, believed to have been formed by a protostellar merger \citep{Bally2020}.
The expansion of the fingers as revealed by proper motions are critical to better understanding the evolution of these structures and TEMPO will offer a new epoch to the community. 

\subsubsection{Investigating Young Stellar Variability}
Young stellar objects (YSOs) exhibit wide-ranging photometric variability, which includes short-term near- or mid-IR fluctuations in brightness.
Prior IR investigations of the Orion region include the \textit{Spitzer} GO6 Exploration Science program known as YSOVAR (Young Stellar Object Variability), which provided mid-IR (3.6 and 4.5\,$\mathrm{\mu}$m) photometric variability survey of YSOs in a 0.9 square degree region centered on the Trapezium cluster \citep{morales-calderon2012}. This survey generated light curves for over more than 1,000 YSOs with a cadence of 2 epochs per day for 40 days. 

This cadence enabled precise measurements of YSO spin periods driven by star-spot modulation, which are generally are on the order of 2-10 days.
However, shorter cadence data is required for investigations of flares.
YSOs are active flare stars with strong magnetic fields and X-ray emission with flare durations ranging between minutes to hours, making short cadence investigations incredibly useful.   
TEMPO will provide unprecedented short cadence investigations, high photometric precision, and a long temporal baseline.
As such, TEMPO will be sensitive to detecting and characterizing continuum, Paschen-alpha, and Paschen-beta flares.   
Incorporating the Roman F213 filter for near-simultaneous imaging will permit the ability to distinguish continuum flares from those dominated by H-recombination lines.  

\citealt{2019ApJ...872..183F} combined \textit{Spitzer} and WISE observations to search for YSO outbursts, constraining their occurrence rate in the ONC.
The team found that such events are far more common at earlier stages of evolution, occurring every $10^{3}-10^{4}$\,yr. They also highlighted the ability of Roman (then referred to as WFIRST) to contribute to further analysis of eruptive YSOs.
Given the fact that thousands of YSOs will be present in the TEMPO field of view, it is feasible to search for far more luminous, but rare, accretion bursts similar to those witnessed in McNeil's Nebula, or akin to EX Ori and FU Ori outbursts \citep[e.g.,][]{1996ARA&A..34..207H,2004ApJ...606L.123B,2009ApJ...692L..67A}.
Further, many embedded YSOs also produce compact, near-IR reflection nebulae, whose illumination varies on time-scales ranging from hours to weeks, thereby also requiring short cadence observations.
The time scale in this case correspond with the orbital periods of clumps within the accretions disks \citep[e.g.,][]{Muzerolle2013,Balog2014}.  
It will also be possible to leverage observational synergies with the Vera C.~Rubin Observatory Legacy Survey of Space and Time (hereafter Rubin). 
More specifically, Rubin will offer complementary deep, wide-field, multi-band optical observations at a several-day cadence over an expected decade-long baseline \citep{2022arXiv220212311G}. 
Rubin will sample both short-term (hours to days) and long-term (months to years) variability, identifying more YSOs than ever before \citep{Bonito2023}.
Yet, it will be difficult to pinpoint the specific star responsible for these signatures without a synergy such as Roman's spatially-resolved NIR observations.
Further, Roman data would also offer the opportunity to further characterize the target when it is active and/or quiescent.

A particularly important class of YSOs are dipper stars --- young stars that exhibit photometric dips in brightness lasting 0.5-2\,days with depths of up to 50\%. 
Nearly one thousand dipper stars have been observed to date \citep[e.g.,][]{Ansdell2016,Rodriguez2017,Ansdell2018,Hedges2018,Bredall2020, 2022ApJS..263...14C,2023MNRAS.521.1700M}.
One possible mechanism driving the brightness variations is the obscuration and accretion of material in the surrounding circumstellar disk, orbiting near the co-rotation radius \citep[e.g.,][]{Ansdell2016,Bodman2017}.
If this suggested underlying mechanism is truly at play, the photometric dips observed by the Roman F146 IR spectral band will be statistically weaker than prior dipper detections performed in the optical, as NIR investigations penetrate through dust and gas much more effectively than visible light.
Therefore, TEMPO provides a valuable opportunity to test this theory, while also characterizing the variability using short cadence integration.
Furthermore, the ONC is an ideal location to test such a theory, as the ages of the stars is this region coincides with the peak age of the dipper star age distribution, as determined by \citet{2022ApJS..263...14C}. 
Furthermore, the ONC is already well known to be rich with test cases. 
For example, a three-year Submillimeter Array (SMA) survey found 42 protoplanetary disks (0.003-0.07\,\Msun{}) in the ONC \citep{Mann2010}. 
While signs of photoevaporation were observed in the outer regions of the observed protoplanetary disks, the team found that 18\% of these disks had masses greater than 0.01\,\Msun{} within 60\,AU. 

One relatively novel class of young variable star is the complex periodic variable, which are understood based on the sharpness and duration of their variability to be explained by clumps of circumstellar material located at the stellar corotation radius \citep[e.g.,][]{Stauffer2017,Bouma2023}.  The origin of the material is not known, but it is likely composed of either gas or dust.  The approximately 100 known instances of these stars are all young ($<200$\,Myr), rapidly-rotating ($P\lesssim 2$\,d), low-mass ($M<0.4$\,\Msun{}) M dwarfs \citep[e.g.,][]{Rebull2016,Stauffer2017,Stauffer2018,Zhan2019,Gunther2022,Bouma2023, Popinchalk2023}.  They are distinct from dippers in that they can be a factor of ten older, and they do not show any infrared excess, imposing a limit on the amount of warm dust that can be present near the star.  Spectroscopically, they are weak-lined T Tauris, indicating no evidence for active accretion.  The light curves themselves are also periodic over hundreds of cycles, rather than stochastic over just a few cycles \citep{Bouma2023}.

In their survey of the $\sim$3\,Myr Taurus association, \citet{Rebull2020} found that 3\% of Taurus members show highly structured and periodic optical light curves.  The TEMPO survey plans to observe roughly 2,000 young stars (plus an additional 5,000 stars in the field).  We therefore expect to discover 60 complex periodic variables, if the occurrence rate from Taurus translates to Orion.  This will be important for two reasons.  First, the lower mass limit for CPVs is not currently known.  The lowest stellar masses currently known to exhibit this phenomenon are $\approx$0.12$M_\odot$ \citep{Bouma2023}.  The TEMPO survey will extend our sensitivity to lower-mass objects and, in particular, show whether brown dwarfs can be CPVs.  Second, a larger CPV sample will clarify a potentially important source of false positives for transiting exoplanets.  
PTFO 8-8695, a candidate 7--10 Myr hot Jupiter in Orion \citep{vanEyken2012}, has been argued to be a binary system in which one component is a complex periodic variable \citep{Bouma2020}.  If true, the saga of that object's discovery and follow-up suggests that care in light curve classification will be paramount, particularly for candidate exoplanets with orbital periods near their host star's rotation period.

There are also important filter considerations in this subdomain.
In addition to the ONC, the OMC1, OMC2, and OMC3 clumps in the ISF have been shown to host hundreds of embedded YSOs. While most are not visible in the optical bands, they many do appear on 2-micron K-band surveys.   
The F213 filter will characterize these YSOs better than the F146 filter, allowing the TEMPO survey to investigate a younger sub-population of forming stars and planetary systems than the ONC.   

\subsubsection{Stellar Multiplicity: Eclipsing Binaries and Stellar Companions}
TEMPO will yield insight into binary stars on multiple fronts. 
There are $>40$ known double-lined spectroscopic binaries (SB2s) within the ONC \citep{kounkel2019}. 
Although no eclipses have been detected among these systems to date with surveys such as \tess{}, higher resolution and higher sensitivity data from TEMPO may change that. Additionally, independent detection of eclipsing binaries \citep[such as six systems detected in \textit{Spitzer} light curves,][]{morales-calderon2012} will also allow for subsequent radial velocity follow up observations of these systems. The identification of young eclipsing SB2s will allow for direct dynamical determination of both masses and radii of these stars, which is a prerequisite for testing pre-main sequence evolutionary models.

Furthermore, given a resolution of 0.105'' in the F146 band (point spread function at the full width at half maximum), it will be possible to resolve widely separated binaries down to separations of $<30$\,AU \citep{de-furio2019}, which provides a crucial test of universality of the binary function in young clusters, and the role of the environment on binary formation. 
ONC, and other young star-forming regions, have previously been suggested to have multiplicity fraction highly discrepant at these separations relative to the field \citep{duchene2018, defurio2021}, particularly for lower mass stars. 
Moreover, it has been noted that dissimilar binary populations may be present in dense high-mass clusters like the ONC \citep{Tokovinin2020}. 
However, as the current census of stars in the ONC that can be resolved down to these separations is still low, the larger sample that TEMPO will offer will yield tighter constraints on the multiplicity fraction that could allow to more definitively examine the evolution of currently young binaries with respect to the field.

Binary star formation remains an unsolved problem. \citealt{Raghavan2010} performed the often used definitive study of stellar binarity in solar-age field stars. These authors found that the mass ratio distribution peaks near $M_2/M_1=q=1$ and is flat to smaller values with the binaries having a mean separation of 50\,AU. However, binary stars harboring planets seem not to follow this paradigm having mean separations near 100\,AU and $q$ values which seem flat \citep{Ziegler2020, Lester2021,Howell2021A}. 
Solar-like stars (F,G,K types) have a binary fraction near 50\%, while M type stars have lower a binary fraction, near 25\% \citep{Winters2019,Offner2023}. 
Field M star binaries tend to favor higher $q$ values and show a mean separation near 20\,AU \citep{Winters2019}, while those hosting exoplanets show a flat or increasing number toward smaller $q$ values and a mean separation near 60\,AU (Matson et al., 2024, submitted). 
These differences in field binaries with no detected planets versus the binaries with detected planets are being attributed to planet formation and planetary orbital evolution and dynamics within the star systems, but the exact mechanisms at work are yet to be discovered. 
An analysis of spatial correlations is warranted. For example, if disk truncation is dominated by photo-evaporation from the massive stars in the ONC, we would expect to see mores instances of this effect at closer distances to the Trapezium stars. 
If, on the other hand, dynamics are the driving mechanism, we would expect to see a correlation with stellar number density. 

A study of both young stars and sub-stellar mass objects in Orion with the contemporaneous study of surrounding field stars will provide ideal and uniform comparison samples between early formation time binaries/planetary systems and old, solar-age field stars.  TEMPO will collect the first large low-mass/sub-stellar sample by which we can explore when and how planets form in binary systems and when and how they affect the mass ratios of the binary and their orbital separation.

\subsection{TEMPO Outcomes on Galactic and Extragalactic Scales}
\label{subsec:extragalactic}    

\subsubsection{Dust Analysis of Orion}
\label{subsec:dust}    
The TEMPO survey, in conjunction with simultaneous ancillary multi-band imaging, will facilitate the generation of an ensemble of NIR extinction curves, charting the evolution of grain size and composition in Orion as a function of column density. 
These data will help unveil how the grain size distribution in a protostellar cloud is connected to that of dust in a more diffuse media.
Indeed, there is evidence in Orion for a systematically different extinction law in regions most affected by stellar feedback \citep{Meingast2018}.
Detailed stellar modeling coupled with the known distance to Orion allows the absolute extinction law to be measured. 
The true steepness of the NIR extinction law remains a major uncertainty in the Galactic extinction curve \citep{MaizApellaniz2020, Hensley2021, Decleir2022, Butler:2023}. 
Orion, with its wide range of environmental conditions and proximity, provides an ideal laboratory to understand it.

Switching between Roman filters is beneficial in this subdomain, as it enables a pixel-by-pixel determination of extinction. Modulation of the light curves by occulting structures in circumstellar environments is likely dominated by dust in the inner disks and envelopes.  A near simultaneous measurement of colors using F146 + F213 filters will distinguish between small particles (particle size $<$ the wavelength in the Rayleigh and Mie regimes) and large particles (particle size $>$ the wavelength in the shadowing regime\footnote{The regime when the grains are much larger than the wavelength so that grains cast shadows.}). 
The non-variable component of extinction will trace the properties of the foreground dust.
%AR: Might also be worth adding how extinction depends on radiation energy density in the ONC? Is it decreased or enhanced near the massive stars? Dust grains and PAHs will be photoevaporated near or within the HII regions.

\subsubsection{Investigating the Nature and Origin of the Radcliffe Wave}
\label{subsec:radcliffe}
The Radcliffe Wave (RW) is a recently discovered coherent structure in the disk of the Milky Way \citep{Alves2020}. 
The wave is 2.7\,kpc long and is believed to have been formed by young star-forming regions in the solar neighborhood.
The molecular clouds in this region align themselves to form a damped sinusoidal wave-like structure with a maximum vertical displacement of approximately 160\,pc.
The Orion star-forming region is part of this wave structure, however, the nature and origin of this feature found in the dense gas remains elusive. 

Recent work used the proper motions of upper Main Sequence (UMS) stars from the \textit{Gaia} mission \citep{Poggio2021,Zari2021} to study whether these stars are part of the vertical oscillation \citep{2022A&A...664L..13S}.
By tracing the young stellar counterpart of the RW kinematics, it was discovered that the spatial oscillation is accompanied by a wave-like motion in velocity space, resembling a harmonic oscillator with vertical motions extending well beyond the RW itself \citep{2022A&A...660L..12T}. 
In particular, the data showed a small-scale vertical mode in the kinematics of young stars, with the amplitude of the oscillation correlating with the age of the stars. 
However, the dust extinction significantly limited this analysis.

The IR data from TEMPO is needed to measure the kinematics of stars in the RW, which are not observable in the optical band. 
These measurements, combined with the radial velocity and chemical abundances of the data inferred from SDSS-V in the
the same region, will help to determine whether the RW is a bending wave induced by the passage of a satellite galaxy or a feature created by gas instability.

\subsubsection{Detection of Extragalactic Sources}
\label{subsec:detectextraglactic}
TEMPO's observations will achieve a faintness limit on par with that of the Hubble eXtreme Deep Field, which reached a typical depth of $m_{\mathrm{AB}}=30$\,mag \citep{Illingworth2013}. 
More specifically, TEMPO's proposed FOV offers observations with very little extinction in two out of the 18 H4RG photodiode arrays that compose the Roman WFI detector focal plane. 
In addition, at +5~hr right ascension, there are few other extragalactic deep fields. Therefore, with a limiting mag of 29.7\,\ab{}, TEMPO is capable of discovering extragalactic sources that will be ideal for follow-up with forthcoming 20-30 meter class telescopes.

\section{Conclusion}
\label{sec:conclusion}
In this paper, we have outlined the ways in which the TEMPO survey has the potential to address a wide range of scientific open questions with a single, coherent dataset, thereby enabling scientific investigations across multiple astrophysical subdomains.
The survey will enable observations with deep near-infrared magnitude limits, wide field coverage, high cadence, and a long temporal baseline, producing a dataset of lasting scientific value.
The key science drivers addressed by this program include the ability to: constrain prevailing theories of the formation and evolution of stars, planets, and moons; discover widely-separated exoplanets to better under the census demographics of these systems; investigate the variability of young stars and brown dwarfs; trace the evolution of stellar jets and outflows; and constrain the low-mass end of the stellar IMF.

The duration of the TEMPO survey is the equivalent to 475 HST orbits, a baseline that cannot be requested through the standard Telescope Allocation Committee calls. 
However, TEMPO will be well poised to be an impactful Roman Treasury Program. 
TEMPO's deep coverage and broad band imaging enables the detection of objects one hundred times fainter than prior treasury programs in the ONC \citep{Limbach2023}.
In addition to exploring new discovery spaces, TEMPO will build upon prior efforts, including \textit{The HST survey of the Orion Nebula Cluster} (GO10246, 104 orbits; \citealt{Robberto2013}) and \textit{The Orion Nebula Cluster as a Paradigm of Star Formation} (GO13826, 52 orbits; \citealt{Robberto2020}) --- both Hubble Space Telescope Treasury Programs.
Furthermore, Roman attains HST survey depth in significantly less time. To illustrate, while HST required approximately 35 hours to map a 0.135 square degree area, Roman achieves equivalent photometric precision in a field double the size within 45 minutes. A 30-day Roman survey will be comparable to conducting an HST survey nearly 1000 times.
Pushing the boundaries of efficiency and depth, TEMPO is poised to leave a lasting legacy in our exploration of the cosmos.

\begin{acknowledgments}
The authors will like to acknowledge helpful discussions with Morgan MacLeod, Snežana Stanimirović, and Anne Noer Kolborg.
MSF gratefully acknowledges the generous support provided by NASA through Hubble Fellowship grant HST-HF2-51493.001-A awarded by the Space Telescope Science Institute, which is operated by the Association of Universities for Research in Astronomy, In., for NASA, under the contract NAS 5-26555.
This work was carried out in part at the Jet Propulsion Laboratory, California Institute of Technology, under a contract with the National Aeronautics and Space Administration.
RY is grateful for support from a Doctoral Fellowship from the University of California Institute for Mexico and the United States (UCMEXUS) and CONACyT, a Texas Advanced Computing Center (TACC) Frontera Computational Science Fellowship, and a NASA FINESST award (21-ASTRO21-0068). J.~M.~V. acknowledges support from a Royal Society - Science Foundation Ireland University Research Fellowship (URF$\backslash$1$\backslash$221932).  
R.~H. acknowledges support from the German Aerospace Agency (Deutsches Zentrum f\"ur Luft- und Raumfahrt) under PLATO Data Center grant 50OO1501.
\end{acknowledgments}

\bibliographystyle{aasjournal}
\bibliography{bibliography}
\end{document}

%% file: macros.tex
\newcommand{\tess}{\textit{TESS}}
\newcommand\Msun{$\rm M_{\odot}$}

\newcommand\MJ{$\rm M_{\mathrm{J}}$}
\newcommand\RJ{$\rm R_{\mathrm{J}}$}
\newcommand\RE{$\rm R_{\Earth}$}
\newcommand\ME{$\rm M_{\oplus}$}
\newcommand\ab{$\mathrm{mag_{AB}}$}